# A progressively reduced pretension method to fabricate Bradbury-Nielsen gates with uniform tension.

Kai Ni[1,a)], Jingran Guo[1], Zhou Yu[1], Like Cao[1], Quan Yu[1], Xiang Qian[1], Xiaohao Wang[1,2]

[1] *Division of Advanced Manufacturing, Graduate School at Shenzhen, Tsinghua University, Shenzhen, 518055, China*

[2] *State Key Laboratory of Precision Measure Technology and Instruments, Tsinghua University, Beijing 100084, China*

[a)] E-mail:ni.kai@sz.tsinghua.edu.cn

A Bradbury-Nielsen gate (BNG) is often used to modulate ion beams. It consists of two interleaved and electrically isolated sets of wires with uniform tension, which can keep parallel, equidistant, and coplanar over a wide temperature range, making the BNG reliable and robust. We have previously analyzed the non-uniform problem of wire tensions with sequentially winding method, and developed a template-based transfer method to solve this problem. In this paper, we introduced a progressively reduced pretension method, which allows directly and sequentially winding wires on the substrate without using a template. Theoretical analysis shows that by applying proper pretension to each wire when fixing it onto the substrate, the final wire tensions of all wires can be uniform. The algorithm and flowchart to calculate the pretension sequence are given, and the fabrication process is introduced in detail. Pretensions are generated by weights combination with a homebuilt weaving device. A BNG with stainless steel wire and a printed circuit board substrate is constructed with this method. The non-uniformity of the final wire tensions is less than 2.5% in theory. The BNG is successfully employed in our homemade ion mobility spectrometer, and the measured resolution is 33.5 at a gate opening time of 350 μs. Compared to the template-based method, this method is simpler, faster and more flexible when making BNGs with different configurations.

**I. INTRODUCTION**

Ion gates are widely used in analytical instruments such as ion mobility spectrometer (IMS)[1-5] and time-of-flight mass spectrometer (TOF-MS)[6-8] to modulate the beams of charged particles. The most convenient and classical form of an ion gate is the Bradbury-Nielsen gate (BNG)[9-14], consisting of two electrically isolated sets of parallel and equally spaced wires in the same plane. The BNG can modulate ion beams quickly and efficiently by periodically loading transverse electric fields in the adjacent wires. Figure 1 shows the working mode of a BNG in IMS. When the two interleaved wire sets are at the same voltage (open state), the charged particles can pass through the BNG and their moving direction is barely affected. When a potential difference is applied to the adjacent wires, a transverse electric field is created (closed state), the gate deflects the charged particles away from their original moving direction and causes them to neutralize on certain electrodes (in IMS) or miss target detectors (in TOF-MS). To make sure the BNG working reliably with high-performance, the adjacent wires in a BNG should be electrically isolated, parallel, equidistant, and coplanar to obtain a uniform transverse electric field in the closed state, which helps the BNG achieving the highest stop efficiency and requires the minimum voltage difference. Therefore, the BNG fabrication method should realize the goal of precise positioning and electrical isolation. Besides, easy operation and time-efficient are also important.

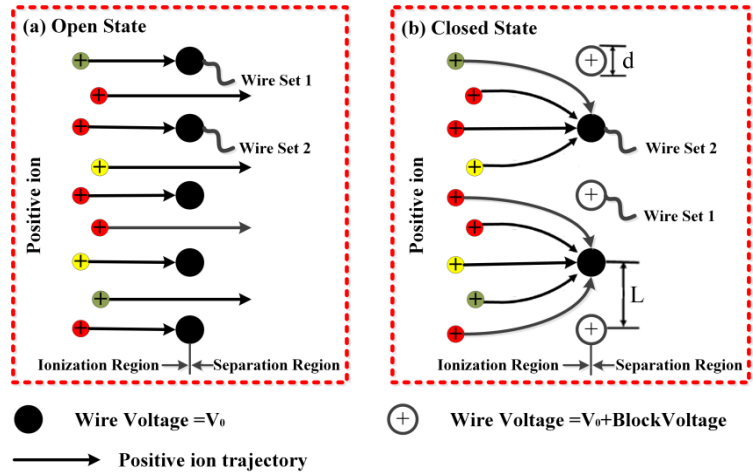

FIG 1. The operation mode of a BNG in IMS. (a) When the two interleaved wire sets are at the same voltage, ions can pass through the gate. (b) When a block voltage is applied to one wire set, all ions will neutralize on the wire set with lower potential.

Over the past decades, researchers have developed various methods to fabricate BNGs, such as hand-weaving methods including direct manual winding[15-20] and template-based transfering[21-22], micromachining methods including bulk silicon micromachining[23], deep reactive ion etching (DRIE)[24] and photo-etching[25]. As the micromachining methods require clean room and are complex and time-consuming, they are not widely used in the BNG fabrication. But they represent a substantial upgrade for applications where microfabrication capabilities are available and the ion beams are smaller than sub-millimeters. Hand-weaving methods are the earliest and most widely adopted methods for fabricating BNGs. The key feature is winding wires on the substrate sequentially with a constant tension. Generally, metal wires are adopted as conductive wires. Polymers such as polyimide, polytetrafluoroethylene, and nylon are often used as insulating substrates. In the direct manual winding method, the metal wires are sequentially wound and fixed directly on the insulating substrate. However, due to the Young's modulus of the wires and the substrate differ significantly, this would easily result in the non-uniformity of the final wire tensions[22]. As BNGs often work in a wide temperature range from 25 ℃ to 150 ℃, causing considerably different deformations between the wires and the substrate, the non-uniformity of wire tensions is altered with temperature in terms of parallelism, spacing, and co-planarity (some wires may even be broken or relaxed). In our previous work, we have reported a simple template-based transfer method to fabricate BNG with printed circuit board (PCB) substrate[22]. The metal wire is first wound around a machined metal template, then the wire mesh is wholly transferred to a PCB substrate. As the template is much stronger than the metal wire, it is much easier to assure the wires wound on the template with uniform tension after the first step, and the following transfer step will maintain the uniformity. High quality BNGs are made with this method compared to traditional sequentially winding method, and the long-time performance is reliable in our homemade IMS. However, the use of template during fabrication is still a little complex. Moreover, one single wire break will cause re-doing the whole work when weaving wires on the template, which seriously decreases the fabrication efficiency.

In this paper, we introduced a progressively reduced pretension method to fabricate BNG with uniform tension. Wires are directly wound on the PCB substrate sequentially without using a template. A homebuilt device is developed to apply proper pretension to each wire during fabrication. A PCB-based BNG is made by this method and its performance is evaluated in IMS.

## II. ANALYSIS AND ALGORITHM

In BNG, Polymers are often used as insulating substrates, and metal wires are used as conductive wires. Both of them are elastic materials. According to the Hooker's law of elastic material under linear response, we can obtain the relationship of object's deformation $\Delta l$ and tension $F$ as expressed in Eq. (1):

$$\Delta l = \frac{Fl}{EA} \tag{1}$$

where $l$ is the original length of the object, $E$ is the object's Young's modulus and $A$ is the object's cross-sectional area. We can see that under the same tension $F$, $\Delta l/l$ is inversely proportional to $EA$. While the substrate's $EA$ is much smaller than the wire's, the strain of the substrate will be much larger than the wire's when they reach force balance. In our previous work[22], we have demonstrated that sequentially fixing wire sections on the substrate with a constant pretension inevitably results in the non-uniformity of final wire tensions. And by using a template with much larger $EA$, the problem can be well solved. In the following analysis, we will show that by applying proper pretension to each wire section, uniform final wire tensions can also be achieved without a template.

Assume pretension $F_P^{(1)}$ is applied to the first wire. After it is fixed on the substrate, both of them will contract to reach a static force balance. We use $F_T^{(1)}$ to represent the wire tension in the balanced state. When fixing the second wire, if its pretension $F_P^{(2)}$ is set equal to $F_T^{(1)}$, i.e. $F_P^{(2)} = F_T^{(1)}$, it is easy to prove that when the two wires and the substrate reach a new balanced state, the two wire tensions will be the same $F_T^{(2)}$. And for the third wire, if $F_P^{(3)} = F_T^{(2)}$, their balanced wire tensions will all be $F_T^{(3)}$. According to this rule, as long as the pretension for the $i$th wire is set equal to the balanced wire tension $F_T^{(i-1)}$ of the previous $i$-1 wires, i.e. $F_P^{(i)} = F_T^{(i-1)}$, when the system is rebalanced, the $i$ wire tensions will all be the same $F_T^{(i)}$. This is exactly what we need to implement a wire grid with uniform tension.

According to the above discussion, when $i$-1 wires have been fixed on the substrate, their balanced wire tensions are all $F_T^{(i-1)}$, and their stretched lengths are the same. This means all wires actually have the same original length $H_0$. Their stretched length $H_0+\Delta H$ equal to the contracted substrate's length $L_0-\Delta L$, where $L_0$ is the original length of the substrate, and $\Delta L$ is the substrate's compression when $(i-1)F_T^{(i-1)}$ applies to it. This can be expressed as in Eq. (2):

$$H_0 \left[ 1 + \frac{F_T^{(i-1)}}{E_1 A_1} \right] = L_0 \left[ 1 - \frac{(i-1) F_T^{(i-1)}}{E_2 A_2} \right] \tag{2}$$

where $E_1$ and $E_2$ is the Young's modulus of the wire and the substrate respectively; $A_1$ and $A_2$ is the cross-sectional area of the wire and the substrate respectively. Then we can get pretension $F_P^{(i)}$ for the $i$th wire

$$F_P^{(i)} = F_T^{(i-1)} = \frac{L_0 - H_0}{(i-1)(L_0/E_2 A_2) + (H_0/E_1 A_1)} \tag{3}$$

Note that $H_0$ is unknown in Eq. (3), and it can be calculated as following: If we want to make a BNG with totally $n$ wires, and the expected final wire tension for all wires is $F_T^{(n)}$, then $H_0$ can be calculated by Eq. (4):

$$H_0 \left[ 1 + \frac{F_T^{(n)}}{E_1 A_1} \right] = L_0 \left[ 1 - \frac{n F_T^{(n)}}{E_2 A_2} \right] \tag{4}$$

By substituting Eq. (4) into Eq. (3), we get

$$F_P^{(i)} = \frac{(n E_1 A_1 + E_2 A_2) F_T^{(n)}}{(i-1)\left[ E_1 A_1 + F_T^{(n)} \right] + \left[ E_2 A_2 - n F_T^{(n)} \right]} \tag{5}$$

With Eq. (5), given $n$ and $F_T^{(n)}$ as input parameters, the pretension sequence $F_P^{(i)}$, $i = 1,2,3,…,n$, for all the $n$ wires can be easily calculated. Note that there is no $L_0$ in Eq. (5), indicating that the same pretension sequence can be used for BNG of any size as long as the required $n$ and $F_T^{(n)}$ are the same.

Equation (5) gives the general expression for the ideal pretension sequence. But in practice, the resolution of the actual generated pretension $F_A^{(i)}$ is limited. For example, we use weights to apply pretension to the wire with our homebuilt weaving device (introduced in section III). The minimum available weight is a certain quality $m_{min}$, such as 1g, 5g or 10g, causing the pretension resolution to be 0.01N, 0.05N or 0.1N respectively. Other ways to generate pretension also have their own resolution limit. As a result, $F_A^{(i)}$ is just an approximation of the ideal $F_P^{(i)}$, and the final wire tensions will have some certain degree of non-uniformity. Therefore, one more verification process is needed to ensure that the final non-uniformity is acceptable, and if not, higher grade weights should be used to get higher pretension resolution, achieving smaller non-uniformity. Figure 2 gives the flowchart to calculate and adjust the actual pretension sequence.

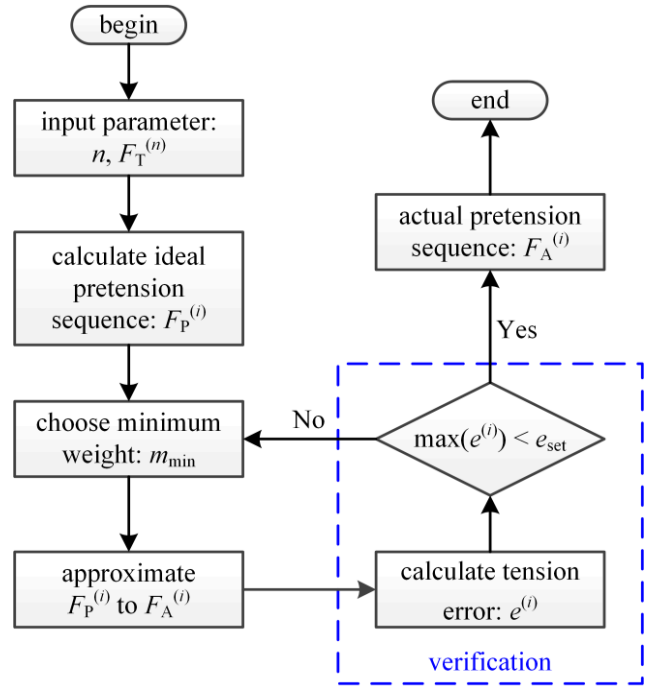

FIG. 2 Flowchart of the calculation of the actual tension sequence.

The verification process is as following: As $F_A^{(i)}$ is different from $F_P^{(i)}$, the wires' original length will no longer be the same, we use $H_0^{(i)}$ to represent the $i$th wire's original length, and $L^{(i)}$ to represent the substrate's length with $i$ wires fixed on it, and $L^{(0)} = L_0$. Then $H_0^{(i)}$ should satisfy Eq. (6) when fixing the $i$th wire:

$$H_0^{(i)}\left[1 + F_A^{(i)}/E_1 A_1\right] = L^{(i-1)} \tag{6}$$

After the $i$th wire is fixed, the static equilibrium between the $i$ wires and the substrate is described as in Eq. (7):

$$\begin{cases} L^{(i)} = H_0^{(j)}\left[1 + F_T^{(i,j)}/E_1 A_1\right], \; j = 1, 2, \ldots, i \\ L^{(i)} = L_0\left[1 - \left(\sum_{j=1}^{i} F_T^{(i,j)}/E_2 A_2\right)\right] \end{cases} \tag{7}$$

where $F_T^{(i,j)}$ is the balanced wire tension of the $j$th ($j = 1,2,\ldots,i$) wire when $i$ wires have been fixed. With Eq. (6) and (7), we can recursively get $H_0^{(1)}$, $L^{(1)}$, $H_0^{(2)}$, $L^{(2)}$, $H_0^{(3)}$, $L^{(3)}$, …, $H_0^{(n)}$, $L^{(n)}$ in order. When all $H_0^{(i)}$ and $L^{(n)}$ are determined, $F_T^{(n,j)}$ can be calculated with Eq. (7), which indicates the final wire tension of the $j$th ($j = 1,2,\ldots,n$) wire. Then the non-uniformity can be measured with Eq. (8)

$$e_{\max} = \max\left|\frac{F_T^{(n,j)}}{\text{mean}\left(F_T^{(n,j)}\right)} - 1\right| \tag{8}$$

If $e_{\max}$ is less than the expected non-uniformity threshold $e_{set}$, the approximated $F_A^{(i)}$ can be used to make the BNG, otherwise, we should select higher grade weights with smaller $m_{\min}$, recalculate $F_A^{(i)}$ and repeat the verification process.

Based on the above analysis, we calculated parameters to make a BNG with steel wire and PCB substrate. The Young's modulus of steel and PCB are $E_1 = 210$ GPa and $E_2 = 2.7$ GPa respectively. The wire diameter is 0.1 mm, so $A_1 = 7.854 \times 10^{-3}$ mm$^2$; the cross-section of the substrate is a rectangle of $A_2 = 24$ mm $\times 1.6$ mm $= 38.4$ mm$^2$; $L_0 = 27$ mm, $F_T^{(n)} = 1$ N, $n = 20$;

the expected non-uniformity is $e_{set} = 3\%$. Different minimum weight quality $m_{min} = $ 1g, 5g and 10g are used to calculate the actual pretension sequence $F_A^{(i)}$ and the final non-uniformity $e_{max}$.

Figure 3 shows the curve of calculated pretensions versus wire index with different $m_{min}$. We can see that the curves monotonically decrease with the increase of $i$, naming this method a progressively reduced pretension method. Furthermore, the slope of the curves also decrease with $i$ increasing, indicating that later fixed wires require higher resolution of pretension, and larger total wire number $n$ will also require higher pretension resolution. Using smaller $m_{min}$ can get more accurate pretensions, but will need higher grade weights which are expensive and need to be more careful to operate. Table 1 gives the calculated final wire tensions with different $m_{min}$. The non-uniformity $e_{max} = 0.56\%$ when $m_{min} = 1$ g and $e_{max} = 2.34\%$ when $m_{min} = 5$ g, while $e_{max} = 4.36\%$ when $m_{min} = 10$ g, which is larger than $e_{set} = 3\%$. Considering the simplification and cost of the fabrication process, we choose $m_{min} = 5$ g.

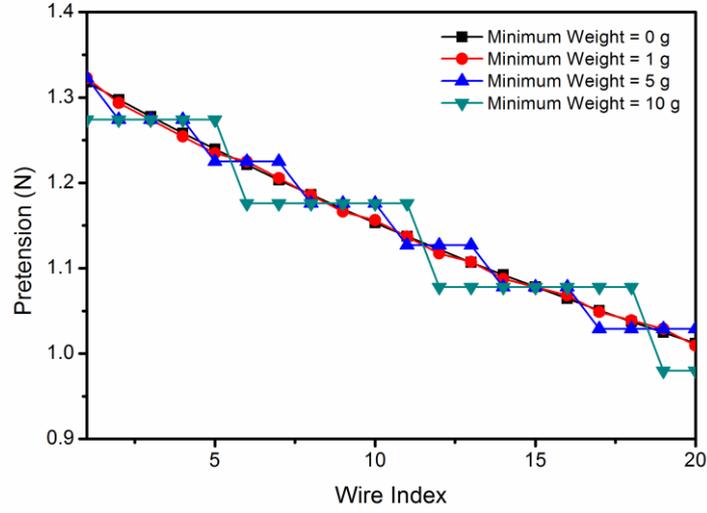

FIG. 3 Plots of pretension sequence with different minimum weight selection.

Table.1. Final wire tensions with different minimum weight

| wire index \ $m_{min}$ | 1 g | 5 g | 10 g |
|---|---|---|---|
| 1 | 1.005 | 1.004 | 0.957 |
| 2 | 0.996 | 0.976 | 0.977 |
| 3 | 0.996 | 0.996 | 0.997 |
| 4 | 0.996 | 1.015 | 1.016 |
| 5 | 0.995 | 0.985 | 1.035 |
| 6 | 1.004 | 1.003 | 0.956 |
| 7 | 1.002 | 1.021 | 0.973 |
| 8 | 1.000 | 0.989 | 0.990 |
| 9 | 0.997 | 1.006 | 1.007 |
| 10 | 1.003 | 1.022 | 1.023 |
| 11 | 1.000 | 0.990 | 1.039 |
| 12 | 0.995 | 1.005 | 0.957 |
| 13 | 1.000 | 1.020 | 0.972 |
| 14 | 0.995 | 0.986 | 0.986 |
| 15 | 1.000 | 1.000 | 1.000 |
| 16 | 1.004 | 1.014 | 1.014 |
| 17 | 0.998 | 0.978 | 1.027 |
| 18 | 1.001 | 0.991 | 1.041 |
| 19 | 1.004 | 1.004 | 0.956 |
| 20 | 0.997 | 1.017 | 0.968 |
| average (N) | 0.999 | 1.001 | 0.995 |
| $e_{max}$ (%) | 0.56 | 2.34 | 4.36 |

## III. FABRICATION METHOD

Wires are positioned precisely by weaving them into the slots of the PCB substrate. The substrate contains equally spaced slots as shown in Fig. 4. The electrical isolation of the two sets of wires is also achieved by the PCB substrate, which has two electrically isolated circuits used to apply control voltage. Each circuit distributes one group of 0.5 mm × 1 mm square pads to solder the wires. The PCB is 1.6 mm thick and is composed of glass fiber-reinforced epoxy resin (FR4). One side of the substrate was bonded with copper foil.

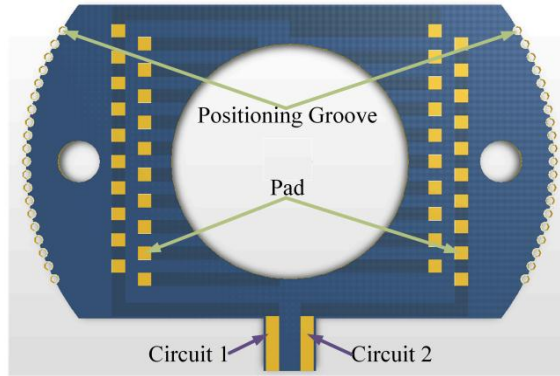

FIG. 4. PCB layout of the BNG substrate.

Controlling the wire pretension is achieved by a homebuilt weaving device, as shown in Fig. 5. The device includes a set of weights, a fixed pulley and a wire spool. The spool consists of two sections, one for providing 0.1 mm diameter stainless steel wire, and the other for winding 0.2 mm nylon wire used to hang the weights through the pulley mechanism. In our design, the two parts of the spool have the same diameter, so the steel wire pretension equals to the weight hang on the nylon wire. Different diameters for the two parts can also be used to generate a scaled pretension by the weights. This scheme can be used to improve the pretension resolution. For example, if the minimum weight is 1 g, a diameter ratio of 1/2 can get a pretension resolution of 0.5 g.

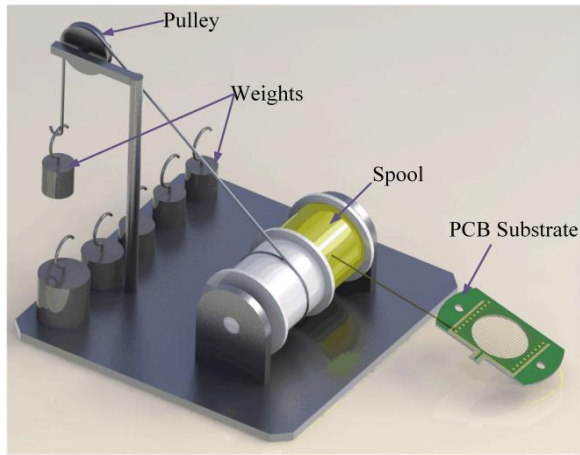

FIG. 5. The homebuilt weaving device.

The fabrication process contains the following seven steps as shown in Fig. 6. Firstly, input the initial parameters to calculate $F_A^{(i)}$ using the flowchart in Fig. 2. And prepare a series of weights which can provide all $F_A^{(i)}$ with combination. Then clean the PCB substrate with acetone before rinsing with deionized water to eliminate possible contaminations. Secondly, use phosphoric acid to corrode the oxide protective layer of the steel wire to facilitate the soldering. Then solder the wire's end onto the first pad. In this process, hang a 5 g weight on the nylon wire to keep the steel wire in tight. Thirdly, replace the 5 g weight to apply $F_A^{(i)}$ to the steel wire through the device. Rotate the PCB substrate manually and guide the steel wire into the positioning slot corresponding to the first pad and locating on the other side of the substrate. Then solder the wire onto the second pad. Fourthly, use side-cutting pliers to cut the wire at the position marked by the red cross. Now the fixing of the $i$th steel wire is completed. Fifthly, if not all wires are fixed, go back to step two, otherwise the fixing process is finished. Finally, the completed BNG is cleaned ultrasonically with acetone and is ready to use. This entire procedure can be completed within 30 minutes. Figure 7 shows an image of a completed BNG, which has a uniform tension, as well as improved parallelism and equidistant spacing.

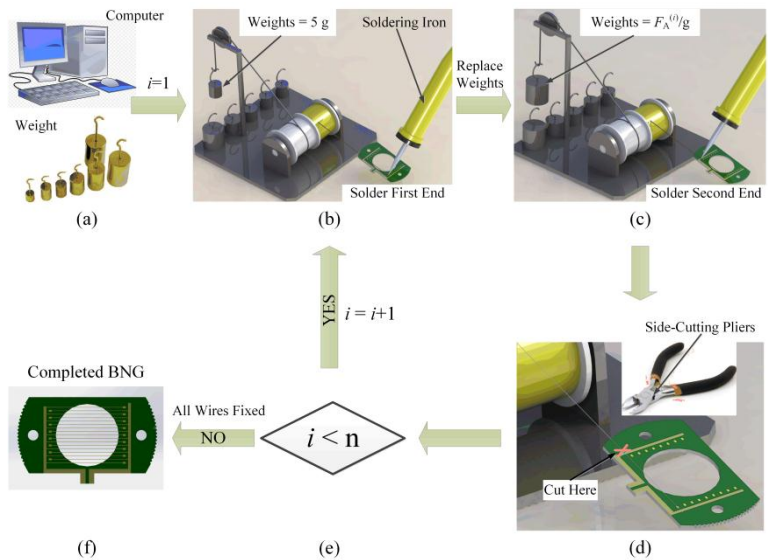

FIG. 6. Schematic of the BNG weaving procedure. (a) Calculate $F_A^{(i)}$ and prepare weights. (b) Solder the wire's first end. (c) Solder the wire's second end. (d) Cut the wire. (e) Determine whether all wires have been fixed. (f) Clean and finish the BNG.

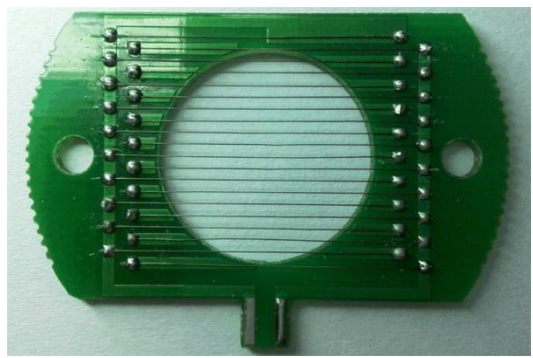

FIG. 7. Photograph of the completed BNG using the progressively reduced pretension method.

## IV. RESULTS

By using this method, we fabricated a BNG with totally 20 wires and 1 N final wire tension. The selected minimum weight is 5 g. The BNG has been installed in a homemade UV-IMS. Experiments were conducted in the UV-IMS to demonstrate the performance of the new BNG. The IMS's drift region length is 81 mm, and its electrical field is 400V/cm. In our instrument, one of the two sets of wires was kept at a constant voltages and the other was switched between high (closed state) and low (open state) voltages (the voltage difference is 100 V). The ion current was amplified by a homemade amplifier with a gain of $10^{10}$V/A. It was then sent to a digital oscilloscope (TDS410A, Tektronix, Wilsonville, OR), which was triggered by the "open" BNG pulse. This pulse served as a start signal for the mobility spectra. These spectra were digitized and averaged by the oscilloscope.

Figure 8 shows the results when analyzing acetone in positive mode. The resolution obtained for the acetone ions was 33.5 at a gate opening time of 350 μs. At the same time, the signal intensity is 0.12 V, which is acceptable comparing to similar reports. In addition, the BNG used for our UV-IMS has been used for over 4 months without any complications or degradation of the materials. And the resolution and sensitivity of the UV-IMS remain unchanged.

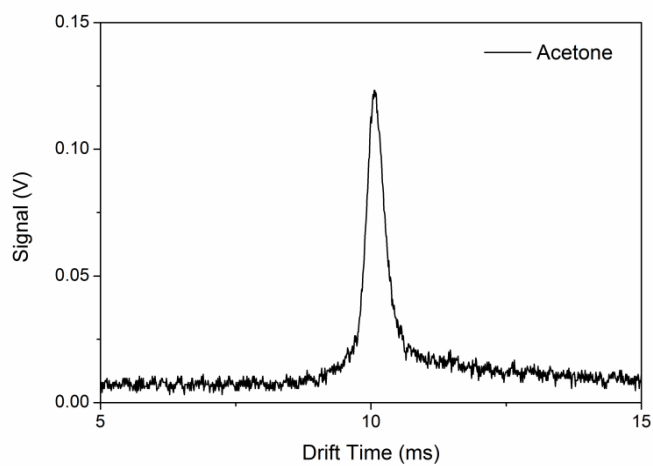

FIG. 8 Acetone spectra obtained on our homebuilt UV-IMS mounted with the new BNG. Drift electric field: 400V/cm; Drift tube temperature: 25 ℃; Drift gas flow rate: 500 sccm; Carrier gas flow rate: 100sccm.

## V. CONCLUSION

We introduced a progressively reduced pretension method to fabricate BNGs with uniform wire tensions. Compared to the previously reported template based method, this method allows directly and sequentially winding wires on the substrate without using a template. The algorithm and flowchart to get the required pretension sequence is given, and theoretical analysis shows that the non-uniformity of the final wire tensions can be less than 2.5% when the weight resolution is 5 g. A PCB-based BNG is constructed with this method, and successfully employed in our homebuilt UV-IMS. The measured resolution of the UV-IMS is 33.5 at a gate opening time of 350 μs. The advantages of this method is simple, fast and flexible. Different BNGs can be made by simply recalculating the pretension sequence instead of refabricating a template. This method can be used to make BNGs for ion mobility spectrometers or time-of-flight mass spectrometers.

## ACKNOWLEDGMENTS

This work is supported by grants from National Natural Science Foundation of China (Grant No. 21205067). We acknowledge Mr. Zhang Xiaoguo for the help of ion mobility spectrometry experiments.

## REFERENCES

[1] M. Kwasnik, J. Caramore, and F. M. Fernández, Anal. Chem. 81, 1587 (2009).

[2] E. J. Davis, K. F. Grows, W. F. Siems, and H. H. Hill Jr, Anal. Chem. 84, 4858 (2012).

[3] G. A. Eiceman, Z. Karpas and H. H. Hill Jr, Ion Mobility Spectrometry, 3nd ed (CRC, Boca Raton, FL, 2013).

[4] X. Zhang, R. Knochenmuss, W. F. Siems, W. Liu , S. Graf, and H. H. Hill Jr, Anal. Chem. 86, 1661(2014).

[5] K. Ni, J. Guo, G. Ou, Y. Lei, and X. Wang, in Proc. SPIE Asia Pacific Remote Sensing, Beijing, China, 14 October-17 October 2014, edited by S. Yang and P. Zhang (International Society for Optics and Photonics, Bellingham, 2014), pp. 925910.

[6] P. G. Hughes, O. Votava, M. B. West, F. Zhang, and S. H. Kable, Anal. Chem. 77, 4448 (2005).

[7] Y. S. Shin, J. H. Moon, and M. S. Kim, Anal. Chem. 80, 9700 (2008).

[8] L. Hua, Q. H. Wu, K. Y. Hou, H. P. Cui, P. Chen, W. G. Wang, and H. Y. Li, Anal. Chem. 83, 5309 (2011).

[9] N. E. Bradbury and R. A. Nielsen, Phys. Rev. 49, 388 (1936).

[10] J. Puton, A. Knap, and B. Siodłowski, Sens. Actuator B, 135, 116(2008).

[11]Y. Du, W. Wang, and H. Li. Anal. Chem. 84, 5700 (2012).

[12]I. V. Kurnin, N. V. Krasnov, S. Y. Semenov, and V. N. Smirnov, Int. J. Ion Mobility Spectrom. 17, 79(2014).

[13]A. T. Kirk, and S. Zimmermann, Int. J. Ion Mobility Spectrom. 17, 131(2014).

[14]K. Ni, J. Guo, G. Ou, X. Zhang, Q. Yu, X. Qian, and X. Wang, Int. J. Mass Spectrom. 379, 75 (2015).

[15]P. R. Vlasak, D. J. Beussman, M. R. Davenport, and G. E. Christie, Rev. Sci. Instrum. 67, 68 (1996).

[16]C. W. Stoermer, S. Gilb, J. Friedrich, D. Schooss, and M. M. Kappes, Rev. Sci. Instrum. 69, 1661 (1998).

[17]A. Brock, N. Rodriguez, and R. N. Zare, Rev. Sci. Instrum. 71, 1306 (2000).

[18]J. R. Kimmel, F. Engelke, and R. N. Zare, Rev. Sci. Instrum. 72, 4354(2001).

[19]A. W. Szumlas, D. A. Rogers, and G. M. Hieftje, Rev. Sci. Instrum. 76, 3 (2005).

[20]Y. Du., H. Cang, W. Wang, F. Han , C. Chen, L. Li, K. Hou, and H. Li. (2011). Rev. Sci. Instrum. 82. 8 (2011).

[21]O. K. Yoon, I. A. Zuleta, M. D. Robbins, G. K. Barbula, and R. N. Zare, J. Am. Soc. Mass Spectrom. 18, 1901 (2007).

[22]K. Ni, J. Guo, G. Ou, Y. Lei, Q. Yu, X. Qian, and X. Wang, Rev. Sci. Instrum. 85 (8), 085107 (2014).

[23]M. Salleras, A. Kalms, A. Krenkow, M. Kessler, J. Goebel, G. Muller, and S. Marco, Sens. Actuator B 118, 338 (2006).

[24]I. A. Zuleta, G. K. Barbula, M. D. Robbins, O. K. Yoon, and Zare, R. N. Anal. Chem. 79, 9160 (2007).

[25]T. Brunner, A. R. Mueller, K. O'Sullivan et al. Int. J. Mass Spectrosc. 309, 97 (2012).